\documentclass{article}
\begin{document}
\title{The Second Covariant Variation of Lagrangian Field Theory.}
\author{Mark D. Roberts, \\\\
Department of Mathematics and Applied Mathematics, \\ 
University of Cape Town,\\
Rondbosch 7701,\\
South Africa\\\\
roberts@cosmology.mth.uct.ac.za} 
\date{\today}
\maketitle
To Appear in the SUSY Encyclop{\ae}dia - Kluwer.\\\\
It has been known,  at least from the time of Synge [1] (1926),
that the second covariant variation of the point particle action
produces the geodesic deviation equation and that this describes the relative
motion of many particles.   The action of the point particle in coordinate
space is 
$S=\int^{\tau_2}_{\tau_1}{\rm d}\tau{\cal{L}}$,
where the Lagrangian is 
${\cal{L}}=-m\ell^n$
and the length is
$\ell\equiv\sqrt{-\dot{x}\circ\dot{x}}$.
Defining the momentum
$P_\nu\equiv\delta S/\delta \dot{x}^\nu$
the first variation of the action gives the equation of motion
$\dot{P}^\nu=0$,
and the second variation gives the geodesic deviation equation
$R^\nu_{.\alpha\beta\gamma}\dot{x}^\alpha\dot{x}^\gamma r^\beta
+\ell(h^\nu_{.\alpha}\dot{r}^\alpha/\ell)=0$,
where $r^\beta$ is the separation vector and 
$h^{\nu\mu}=g^{\nu\mu}+\dot{x}^\nu\dot{x}^\mu/\ell^2$
is the projection tensor.   The equations produced from a second covariant
variation can often be derived from a first order covariant variation of 
a combined action.   
For the point particle a suitable first order combined action is
\begin{equation}
S_c=\int^{\tau_2}_{\tau_1}{\rm d}\tau
\frac{\partial{\cal{L}}}{\partial\dot{x}^\alpha}
\frac{{\rm D}}{{\rm d}\tau}r^\alpha.
\end{equation}
Defining the momenta
$P^\nu\equiv\delta S_c/\delta\dot{r}^\nu$
and
$\Pi^\nu\equiv\delta S_c/\delta\dot{x}^\nu$
variation of the action (1) with respect to $\delta r^\alpha$ gives 
$\dot{P}^\nu=0$ and with respect to $\delta x^\beta$ 
gives the geodesic deviation equation in the form
\begin{equation}
\dot{\Pi}^\nu=R^\nu_{.\alpha\beta\gamma}r^\gamma P^\beta\dot{x}^\alpha,
\end{equation}
substituting in the various definitions gives this equation 
in the familiar from in the text above.

One can implement exactly the same sort of procedure 
for the bosonic string action,  and presumably for more complex actions.
The action of a bosonic string is
$S=\int^{\tau_2}_{\tau_1}{\rm d}\tau\int^\pi_0{\rm d}\sigma{\cal{L}}$,
where the Lagrangian is
${\cal{L}}=-{\cal{A}}/2\pi\alpha'$,
the area is
${\cal{A}}\equiv\sqrt{(\dot{x}\circ\dot{x})^2-\dot{x}^2x'^2}$,
and $\alpha'$ is the string tension.   Defining momenta
$P_{\tau\nu}\equiv\delta S/\delta\dot{x}^\nu$
and
$P_{\sigma\nu}\equiv\delta S/\delta x'^\nu$,
the first variation of the action gives the equation of motion
$\dot{P}^\nu_\tau+P'^\nu_\sigma=0$.
The second variation of the action gives the string deviation equation,
which is a more complex analog of the geodesic deviation equation above.
Again it is simpler to consider the combined action
\begin{equation}
S_c=\int^{\tau_2}_{\tau_1}{\rm d}\tau\int^\pi_0{\rm d}\sigma
\left(\frac{\partial{\cal{L}}}{\partial\dot{x}^\alpha}
      \frac{{\rm D}}{\rm{d}\tau}r^\alpha
     +\frac{\partial{\cal{L}}}{\partial x'^\alpha}
      \frac{{\rm D}}{\rm{d}\sigma}r^\alpha\right).
\end{equation}
Defining the momenta
$P_{\tau\nu}\equiv\delta S_c/\delta\dot{r}^\nu$,
$P_{\sigma\nu}\equiv\delta S_c/\delta r'^\nu$,
$\Pi_{r\nu}\equiv\delta S_c/\dot{x}^\nu$ and
$\Pi_{\sigma\nu}\equiv\delta S_c/\delta x'^\nu$;
first order covariant variation of this action gives 
the string deviation equation in the simple form
\begin{equation}
\dot{\Pi}^\nu_\tau+\Pi'^\mu_\sigma=R^\nu_{.\alpha\beta\gamma}r^\gamma
\left(P^\beta_\tau\dot{x}^\alpha+P^\beta_\sigma x'^\alpha\right).
\end{equation}
Some points to note are:\newline
1)The separation vector $r^\alpha$ is invariant under certain gauge
transformations and in this sense combined actions can be thought of as 
being an unstudied gauge system.\newline
2)A similar analysis can presumably 
be carried out for supersymmetric actions.\newline
3)Third and higher order variations can also be calculated,
there might be a way of evaluating the $n^{th}$ order variation,
and of implementing various calculations on the series.\newline
4)It is not immediate what the best way to quantize such combined actions is,
although this has been done for geodesic deviation,  
Roberts (1996) [2].\newline
5)The implications of the string deviation equation for currently 
understood quantized string theory are not known.\newline
Further details of this work can be found in Roberts (1999) [3].\newline
BIBLIOGRAPHY.\newline
[1]J.L.Synge{\it Proc.Lond.Math.Soc.}{\bf 25}(1926)247.\newline
[2]M.D.Roberts{\it Gen.Rel.Grav.}{\bf 28}(1996)1385.{\tt gr-qc/9903097}\newline
[3]M.D.Roberts{\it Mod.Phys.Lett.}{\bf A14}(1999)1739.{\tt gr-qc/9810043}
\end{document}